\documentclass[12pt]{iopart}
\usepackage{graphics}
\begin{document}

\title[Molecular relaxation paths]{Relaxation paths for single modes of vibrations in isolated molecules}

\author{R Papoular}

\address{Service d'Astrophysique and Service de Chimie Moleculaire

CEA Saclay 91191 Gif-sur-Yvette, France}
\begin{abstract}

A numerical simulation of vibrational excitation of molecules was devised, and used to excite computational models of common molecules into a prescribed, pure, normal vibration mode in the ground electronic state, with varying, controlable energy content. The redistribution of this energy (either non-chaotic or irreversible IVR) within the isolated, free molecule is then followed in time with a view to determining the coupling strength between modes. This work was triggered by the need to predict the general characters of the infrared spectra to be expected from molecules in interstellar space, after being excited by photon absorption or reaction with a radical. It is found that IVR from a pure normal mode is very ``restricted" indeed at energy contents of one mode quantum or so. However, as this is increased, or when the excitation is localized, our  approach allows us to isolate, describe and quantify a number of interesting phenomena, known to chemists and in non-linear mechanics, but difficult to demonstrate experimentally: frequency dragging, mode locking or quenching or, still,  instability near a potential surface crossing, the first step to generalized chaos as the energy content per mode is increased.
\end{abstract}

\maketitle

\section{Introduction}

The work to be described here is an effort to simulate the excitation of pure vibrational modes in a single molecule, then follow their evolution in time, or relaxation paths, with a view to estimating coupling strengths between modes and determining bottlenecks in the way to the ``final" molecular vibration state, that is, before possible subsequent infrared emission, which occurs much later in general.

This effort was triggered precisely by the need to predict, and hence be able to interpret, the observed IR (infrared) emission from isolated interstellar molecules which have been excited by radiation absorption or recombination of radicals (see Williams and Leone \cite{wil}). Such excitation usually leads to electronic state transitions, internal conversion and intersystem crossing, and, at least, mixing of vibrational states. This was masterly described in the pedagogical overview written by Nesbitt and Field \cite{nes}, to which the present work owes much.

 Already in the 80's, Zewail and collaborators \cite{zewa}, \cite{zewb} had illustrated experimentally, and analyzed theoretically irreversible IVR, involving a quasi-continuum of interacting states, but also pointed out the importance for chemistry of non-collisional, non-chaotic, multilevel energy flow, involving only a few coupled states, as evidenced by quantum beats observed in the fluorescence of the excited molecule. They also pointed out the analogy with systems of coupled mechanical oscillators.

At about the same time, it was realized that relaxation in an isolated molecule (non-collisional relaxation) did not always quickly end up in ``thermodynamic equilibrium", as if the molecule were embedded in a thermal bath: rather, it was necessary to acknowledge the existence of mode selectivity (see Sewell \etal \cite{sew}), which may forbid certain regions of molecular phase space and lead to ``restricted" and quasi-periodic flows, as opposed to ``facile", chaotic or downright ergodic motion, which characterizes thermal equilibrium (see Noid \etal \cite{noid}; Brumer \cite{bru}). Chemists were relieved to notice that this need not contradict the celebrated RRKM treatment of uni-molecular reactions: the latter is most generally valid since it does not in effect require complete ergodicity (Sewell at al.\cite{sew}), and since, at sufficiently high excess energy density in the molecule and depending on the extent of overlapping nonlinear resonance interactions (Sibert \etal\cite{sib}, Ford\cite{ford}), ergodicity does set in, as assumed in the first place.

Because of its potential use in selective chemistry, the transition between quasi-periodic and chaotic regimes has since been the subject of considerable interest and numerous experiments (e.g. Sibert \etal\cite{sib}, Sewell \etal\cite{sew}, Deak \etal\cite{deak}, Stromberg \etal\cite{strom}). The experimental approach to these phenomena requires sophisticated apparata, including picosecond spectroscopy and cold molecular beams. Also, for experimental convenience, excited electronic states are privileged over their ground state. Unfortunately, the former lack the desired IR information. Finally, the information about the excited states is mostly deduced indirectly from whatever diagnostic means are available.

Here, we are only interested in the quasi-periodic, or coherent regime, in a limited range of molecules (medium-size hydrocarbons) and take advantage of the recent development of highly performing chemical computational software to simulate the types of molecules we are interested in and their behaviour upon specified excitation into higher vibration levels in the ground electronic state, far from the dissociation limit.

Moreover, the molecular behaviour we are interested in is strongly affected by anharmonocity, and so depends heavily on the energy content of the target excited state. Again, the latter parameter is much  easier to control computationaly than experimentally, as shown below.

Finally, the molecules of astrophysical interest may be composed of as many as a few hundred atoms, and so are very difficult to produce and isolate for excitation and detection, making inevitable the resort to numerical modeling.

For all the imperfections of chemical codes, they are quite convenient (as already noticed long ago: see Noid \etal\cite{noid}), because they provide the possibility of collecting directly the desired information with very high time resolution (10$^{-1}$ fs): in each experiment, all molecular properties are collected as as a function of time, including all bond lengths and orientations, as well as global properties such as kinetic and potential energies, electric dipole moment, etc.

 A key step in carrying out our project was the design of a computational procedure to energize a single normal mode in a chemical model of the real molecular structure, so as to be able to isolate and define clearly individual relaxation paths. This produced sparse spectra and simple beating patterns, from which to deduce coupling energies.

The simulation of relaxation from high energy levels also uncovered chaotic behaviours, difficult to observe experimentally, but well known in non-linear mechanics, when exploring the phase space of non-integrable systems, starting with bifurcations to end up in various forms of chaos. Here are demonstrated, for instance, multiple resonances, localized modes, frequency drag,  mode locking (synchronization) and mode quenching, some of which are seldom treated in the current litterature.

In the following, Sec. 2 is a general reminder of the behaviour of coupled linear oscillators, as an introduction to the formalism, together with the non-linear corrections which were needed to explain the observations described in this work, on common molecules, even at moderate energy contents ($\Delta v=$ 1 or 2). Of course, all of this is best done quantum mechanically (Nesbitt and Field\cite{nes}, Gruebele and Wolynes\cite{gru}, Wu\cite{wu}), but the parallel is so perfect that the classical treatment,
including Classical Dynamics for the motion of nuclei (Born-Oppenheimer approximation), can deliver much of the right physical picture (e.g. Noid \etal\cite{noid}, Kuharski \etal \cite{kuh})(except for quantization and zero-level energy which do not matter here) with only a very light mathematical apparatus (Herzberg\cite{herz}).

Section 3 describes a new procedure for the excitation of single modes and the effects of various parameters defining this procedure. Sections 4 to 9 apply the above to a simple computer-simulated molecule, C$_{2}$H$_{4}$. Several excitation and relaxation routes of interest are considered. Because of the relative sparseness of its vibrational spectrum, it is possible to isolate interesting molecular phenomena and determine some of the coupling factors involved. Section 10 is an extension to a much larger molecule (101 atoms) with a correlatively much denser spectrum.

A general conclusion drawn from these observations is the weakness of most mode couplings and the unequal partition of the excitation energy among the modes at the end of the relaxation path. Only at higher energy contents, the higher the larger the molecule, are chaos and better equipartition, observed (Papoular\cite{pap}), but that is not considered here.

\section{Coupled non-linear oscillators}

In the ground vibrational state the potentials which determine the nuclear motions are parabolic in first approximation. As a consequence, the motion of each atom can be broken down into 3N-6 independent, collective , normal modes, where N is the number of nuclei in the structure.

However, as soon as the molecule is set in motion, deviations from the quadratic dependence of potentials on excursions of nuclei from their equilibrium positions (anharmonicity) changes the normal mode frequencies and couples all these oscillators together, so that energy can flow from one to the other, more or less easily, depending on their coupling coefficients and frquency separations. The ways they do so are the subject of this study.

Consider two oscillators with frequencies $\nu_{1}$ and $\nu_{2}$, coupled by a perturbation term, $v$, in the equilibrium potential. In quantum mechanics, this is an off-diagonal term in the Hamiltonian matrix; in electric circuit theory, it is expressed by the impedance of an electric component common to both coupled circuits. In mechanics, it is the stiffness, $k$, of a spring linking two pendula; for instance, if $m$ is their common mass, the equations of motion are:

$\ddot{x}_{1}+\omega_{1}^{2}x_{1}+\frac{k}{m}(x_{1}-x_{2})=0$

$\ddot{x}_{2}+\omega_{2}^{2}x_{2}-\frac{k}{m}(x_{1}-x_{2})=0$,

where $\omega=2\pi\nu$. In the important particular case of resonance, $\nu_{1}=\nu_{2}=\nu_{0}$, these equations are easily separated into two independent harmonic motions by transforming variables into $X_{a}=x_{1}+x_{2}$ and $X_{b}=x_{1}-x_{2}$, giving

$\ddot{X}_{a}+\omega_{0}^{2}X_{a}=0$

$\ddot{X}_{b}+(\omega_{0}^{2}+\frac{2k}{m})X_{b}=0$

Obviously, the natural frequencies of these so-called normal modes are

\begin{equation}
$$\nu_{a}=\nu_{0}; \nu_{b}=\nu_{0}(1+\frac{2v}{\nu_{0}})^{1/2}, v=\frac{k}{m\omega_{0}}$$.
\end{equation}

One of the new frequencies coincides with the common, unperturbed one, while the other is slightly blue shifted. This corresponds to the quantum classical repulsion of crossing (resonant) energy levels. For sufficiently weak coupling, $\nu_{b}\sim\nu_{0}(1+\frac{v}{\nu_{0}})$. These equations perfectly illustrate, e.g., the symmetric and anti-symmetric scissor modes of C$_{2}$H$_{4}$, which is treated below: $x_{1,2}$ representing the two HCH angles, while $X_{a,b}$ are the two normal modes.

Returning now to the general, out-of-resonance, case, the solutions are of the form

$x_{1}(t)=Acos\omega_{a}t+Bcos\omega_{b}t$

$x_{2}(t)=Ccos\omega_{a}t+Dcos\omega_{b}t$,

where

\begin{equation}
$$\nu_{a,b}=\{\frac{\nu_{1}^{2}+\nu_{2}^2}{2}+v(\nu_{1}\nu_{2})^{1/2}\mp[(\frac{\nu_{1}^{2}-
\nu_{2}^{2}}{2})^{2}+v^{2}\nu_{1}\nu_{2}]^{1/2}\}^{1/2}$$
\end{equation}

and $A, B, C$ and $D$ are constants. Each oscillator is a linear combination of modes $a,b$, with relative amplitudes depending on initial conditions.

The most telling case, then, is when oscillator 1 is initially excited alone. A simple trigonometric transformation of the equations for $x_{1,2}$ reveals that, when the coupling is not too strong, the motion of each oscillator can also be viewed as sinusoidal, with common frequency

\begin{equation}
$$\frac{\nu_{1}+\nu_{2}+{v}}{2}$$
\end{equation}

and amplitude-modulated with frequency

\begin{equation}
$$\Omega=[\frac{(\nu_{1}-\nu_{2})^2}{4}+\frac{v^{2}}{4}]^{1/2}$$.
\end{equation}

 This modulation implies that a corresponding amount of energy is periodically exchanged between the oscillators, as illustrated in fig. 1 (cf. Zewail'z quantum beats). It turns out that the average fraction of energy that is delivered to oscillator 2 is

\begin{equation}
$$\epsilon=[1+(\frac{\nu_{1}-\nu_{2}}{v})^{2}]^{-1}$$.
\end{equation}

\begin{figure}
\resizebox{\hsize}{!}{\includegraphics{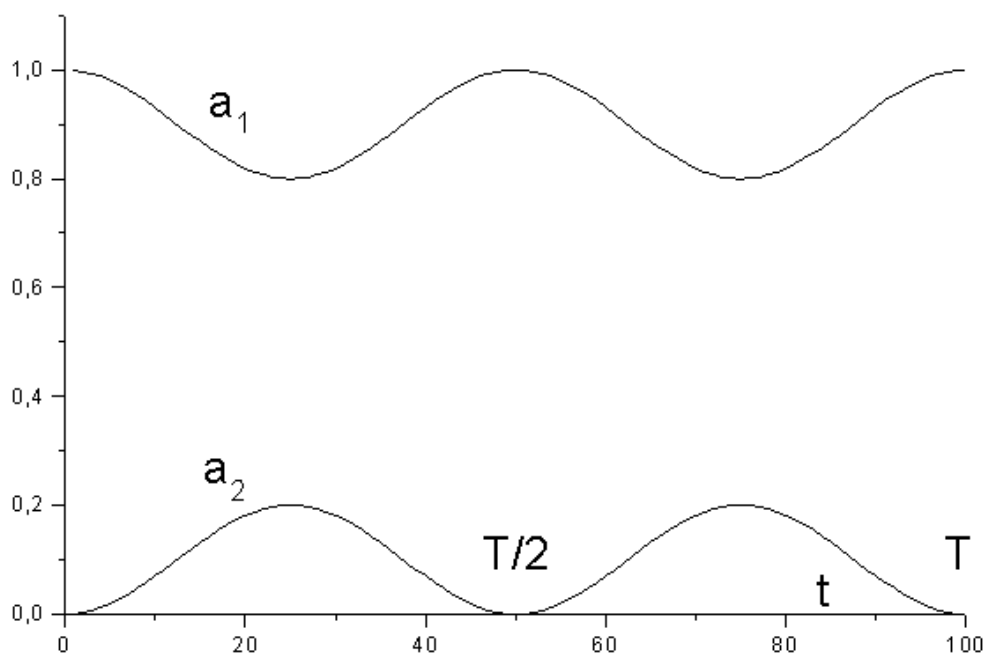}}
\caption[]{Envelopes of the peak amplitudes of excursions $x_{1,2}$  of two coupled oscillators or normal modes, illustrating (4) and (5). Initially mode 1 alone is excited. Both amplitude and periodicity, $T=2\pi/\Omega$, of the curves carry information on both the frequency separation and extent of coupling between the two oscillators.} 
\end{figure}

 Equation (5) predicts that the maximum amount of energy that can be transferred from 1 to 2 increases as the coupling increases and the frequency separation decreases, to reach unity when their ratio becomes infinite (i.e. at resonance). Also, the transfer frequency (4) increases with both parameters. Thus, even in steady state, there is no equipartition in general. In a quantum mechanical context, the squared amplitude of a mode $i$ should be viewed as the probility of finding the system in state $i$ at time $t$; complete transition from state 1 to state 2 ($\epsilon=1$) can only occur in case of resonance  and then only periodically, as in Rabi's oscillations (see Cohen-Tannoudji \etal \cite{coh}). When the coupling becomes too strong, however, the separation into two modes no longer makes sense.

Now, a molecule can be considered as a system of oscillators in different ways. In ethylene, for instance, the two opposite pairs of C-H bonds can be considered as resonant angular oscillators coupled by the C-C bond, as we do in Sec. 4. In a different analogy, when the energy content drives the system out of the parabolic region of the potentials, so the various normal modes determined at low energies can no longer be said to be independent, they can instead be considered as oscillators linked by perturbative couplings. One can then determine the order and extent of such perturbations as a function of energy content. Here, however, the coupling factor also depends on the relative orientations of the vectors describing the nuclear motion in each mode (symmetry selection rules), a constraint which must be taken into account when considering the relaxation ladder and considerably restricts IVR (Felker and Zewail\cite{zewb}). For instance, in the case of a plane molecule such as benzene or ethylene (studied below), out-of-plane CH bends are extremely difficult to excite through the ``planar" modes.

Things are further complicated by the anharmonic dependence of couplings and mode frequencies upon energy content. However, this also gives rise to interesting new phenomena, such as frequency dragging and mode locking, observed below. In order to understand the occurrence of such phenomena, the equations above must be modified to account for deformations of the potential surfaces. Our approach is:

- anharmonicity of $n$th order is approximately simulated by multiplying $\omega_{i}$ by 
$(1-ax_{i}^{n})$, where,as before, $x$ is an oscillation amplitude while $n$ defines the order of anharmonicity; the dimension of $a$ is that of $x^{-n}$ so as to make $ax^{n}$ dimensionless; this implies a dissociation limit at $x_{d}=a^{1/n}$;

-resonances of second order ($\nu_{1}=\nu_{2}+\nu_{3}$, including Fermi resonance as a special case) are approximately simulated by adding to each equation a term $b(x_{1}-x_{2})^2$.

At that point, of course, the equations must be solved numerically for particular values of the parameters. This was done at every step of this work, so as to understand the physical origin of  features exhibited by real molecules and described below. The individual role of each non-linear term is highlighted although they both stem from the same deformation of potential surfaces with increasing energy content and, therefore, are not separable.

\section{Energizing a normal mode}

A particular case of the above is the feeding of energy from an external source (e.g. a single-mode laser) into a single oscillator (a molecular normal mode) of frequency $\omega_{0}$ and initially at rest. Classically, it is well known that, in response to an an applied resonant perturbation, the oscillator amplitude increases as $t$ and its energy as $t^{2}$, which corresponds to $\Omega=0$ in (4). In order to excite one mode of a model molecule in this way, it is in principle necessary to devise a means to perturb all bonds in tune, at the right frequency  and according to the given mode geometry. This is presently almost out of reach in most practical cases because of software limitations. However, we show here that the desired result can be approached by periodically applying short perturbative pulses to a few specific bonds at the same time. If the pulse period is $T=\frac{2\pi}{\omega}$, this is equivalent to a harmonic series of frequencies $\frac{n}{T}$ ($n$ integer $\geq0$). It turns out that, if $\omega$ is slightly smaller than $\omega_{0}$, the procedure nicely takes care of the progressive, anharmonic, decrease of $\omega_{0}$ as energy is fed into it, and that harmonics are not significant.

Let us now quantify these considerations to optimize the result. This is done with the help of fig. 2, where time is in abscissa and bond length (or oscillation amplitude) in ordinate. At times $t=nT$, one or more particular bonds are set to the same (small) value, $q_{0}$, of excursion from equilibrium in length or orientation. Let $\omega_{0}T=2\pi(1-\delta)$. In between two successive pulses at $t=0$ and $t=T$, e.g., the free "motion" of a bond is

$Q_{1}^{t}=q_{1}cos(\omega_{0}t+\phi_{1})$.

After the 2nd pulse, the motion is similarly described by

 $Q_{2}^{\tau}=q_{2}cos(\omega_{0}\tau+\phi_{2})$

with $\tau=t-T$. The procedure imposes

\begin{figure}
\resizebox{\hsize}{!}{\includegraphics{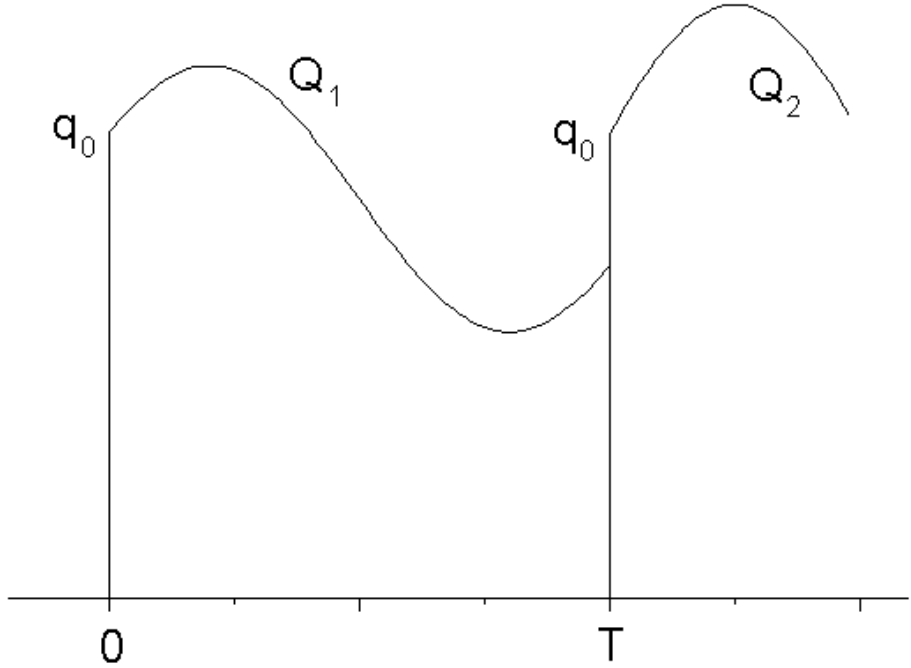}}
\caption[]{One period of the energizing process, illustrating Sec. 3. $Q$ is the free motion of the oscillator that is being energized, between two succeeding, externally imposed, excursions, $q_{0}$.} 
\end{figure}

$Q_{1}(0)=Q_{2}(0)=q_{0}$, 

so there is, at $t=T$, a discontinuity between $Q_{1}(T)=q_{1}cos\phi_{1}$ and $Q_{2}(0)=q_{2}cos\phi_{2}$. However, their derivatives must be continuous, i.e.

$q_{1}sin(\phi_{1}+2\pi(1-\delta))=q_{2}sin\phi{2}$.

Combining these equations gives

\begin{equation}
tan\phi_{2}=\frac{sin[\phi_{1}+2\pi(1-\delta)]}{cos\phi_{1}}=\frac{sin(\phi_{1}-2\pi\delta)}{cos\phi_{1}}.
\end{equation}

Clearly, phase $\phi_{i}$ changes all along, tending to a limit

\begin{equation}
$$\phi_{l}=\pi(\delta-1/2)$, for $\delta\geq0$$,
\end{equation}
\begin{equation}
$$\phi_{l}=\pi(\delta+1/2)$, for $\delta\leq0$$.
\end{equation}

Correlatively, amplitude $q(n)$ tends in time towards a corresponding upper limit 

\begin{equation}
$$q_{l}=\pm\frac{q_{0}}{sin\pi\delta}$$,
\end{equation}

depending on whether $\delta>0$ or$<0$ respectively. In fig. 2, this asymptotic limit (saturation) corresponds to a situation where, because of the increase in amplitude, both ends of the arch of sinusoid have ordinates $q_{0}$. It also appears that $q_{l}\rightarrow\infty$ as $\delta\rightarrow0$, but this has a price in that each pulse then delivers a correspondingly smaller amount of energy to the target mode, thus slowing the process. As $\mid\delta\mid$ increases, the saturation limit is reached earlier but is less energetic.

As the energy to be deposited  may reach a few tens of eV (IR photon energies), anharmonicity cannot be ignored: during the energizing process, $\omega_{0}$ decreases and $\delta$ increases. By way of an illustration, put

\begin{equation}
$$\delta(n)=\delta_{0}+aE_{n}^{m},  \;E_{n}=q_{n}^{2}/2$$.
\end{equation}

Here, $\delta$ is the dimensionless slip in frequency, $n$ is the number of elapsed periods, $m$ is the order of anharmonicity, $E_{n}$ is proportional to the oscillator energy at time $t=nT$ and the dimension of $a$ is such as to make $aE_{n}^{m}$ dimensionless (this convention is slightly different from that of Sec. 2, but carries the same information as to the degree of anharmonicity).

\begin{figure}
\resizebox{\hsize}{!}{\includegraphics{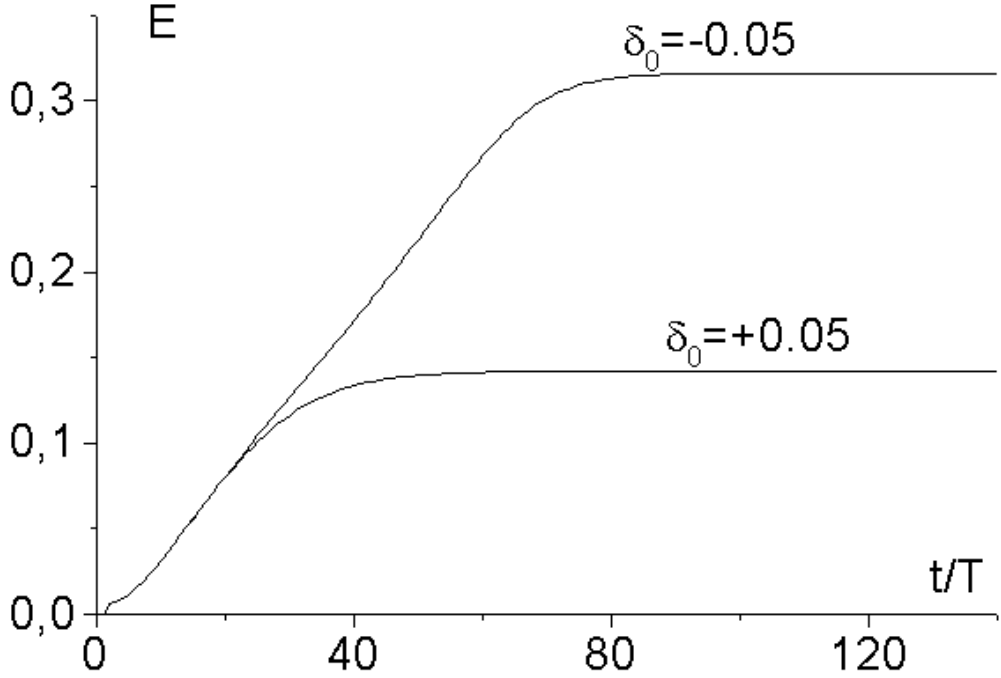}}
\caption[]{Computed examples of energizing curves. Symmetric initial detunings between the source and receiver oscillators yield different limit energies: negative detunings are favored because of anharmonic red-shift. As defined by eq. (10), $E$ is proportional to the oscillator energy at time $t$.} 
\end{figure}

Figure 3 shows the evolutions for two symmetric initial detunings, $\delta_{0}=\pm0.05$, and the same non-zero anharmonicity, $a=0.5$. Negative initial detunings ultimately deliver more energy to the target mode: this is because resonance is approached more closely as the effective detuning decreases in the process; in the limit when the mode synchronizes with the perturbation, its energy tends to 

\begin{equation}
$$E_{l}=(\frac{\mid\delta_{0}\mid}{a})^{1/m}$$,
\end{equation}

if $m$ is constant. In the absence of anharmonicity, of course, both curves coincide, and both limiting deposited energies decrease as the initial detuning increases. The optimum value of the latter is highly sensitive to $a$ and $m$, and must be determined by cut-and-try; but a few trials show that, in general, a few percent negative initial detuning ($\omega_{0}T>2\pi)$ yield enough energy after a few tens of energizing cycles. Thus, in the case of fig. 3, $\delta_{0}=-0.05$ yields a limiting amplitude gain $\frac{q_{l}}{q_{0}}=8$; this corresponds to an energy gain of 64 w.r.t. the potential energy perturbation induced by the initial perturbation, $q_{0}$, of the equilibrium bonds.

As to this initial perturbation, it should match as closely as possible the geometry of the target mode. While this may be easy for small molecules (see Sec. 4 ff), it is hopeless for large ones (Sec. 9). In that case, however, one can select a few bonds of the adequate type, and perturb them in the right way ( stretch, in-plane or out-of-plane). Thus, it was found in this study that an initial C-C stretching perturbation, for instance, is quite helpful in exciting many skeleton and functional modes, albeit with reduced efficiency.

When the periodic perturbation does not perfectly match the desired mode geometry, what is excited is a localized mode (a ``bright" state), i.e. a linear superposition of normal modes, which may blur the relaxation path. However, the frequency selectivity of the energizing procedure is often sufficient to eliminate residual undesirable modes. Note that the situation is quite similar to that of the excitation of modes in a microwave cavity by application of a voltage to a suitably shaped and oriented metallic probe protruding into the cavity.

Following the remark at the beginning of this section, the selectivity of the proposed procedure can be improved by increasing the number of applied perturbations ($q_{0}$) per cycle, for instance including $-q_{0}$ at $t=T/2$, which eliminates all even harmonics.

Once an energizing curve has been "experimentally" obtained, it is possible to determine the number of cycles necessary to deposit exactly the desired number of quanta in the target mode for further study; for instance, to determine the anharmonicity parameters, or mode couplings as a function of energy content.

Note again that, between two succeeding applications of $q_{0}$, the system oscillates freely; the chemical numerical simulation codes used here to implement the whole scheme are Classical Dynamics codes associated with either the semi-empirical code AM1 for small molecules (using quantum mechanics to determine electronic configurations at each step of the dynamics), or the Molecular Mechanics code MM2 (using experimentally adjusted force fields) for large molecules, as embedded in the Hyperchem 5.0 package (Papoular\cite{pap}). An excellent overview of these codes, as well as references to the original papers can be found in Hinchliffe \cite{hin}.

In the following computations, the ambient temperature is 0 and all dynamics are computed by steps of 1 fs.

\section{Case study: C$_{2}$H$_{4}$, coupling energy}

C$_{2}$H$_{4}$ has 12 normal modes between 800 and 3200 cm$^{-1}$. Two of them, the symmetric (sym) and anti-symmetric (a-s) scissor modes, provide a good example of the different ways coupled-oscillators theory can be applied. In the first, both CH bond angles oscillate in phase, and in the second, with opposite phases. These modes can be considered as two oscillators, isolated at low energy contents, which are coupled by anharmonicity at higher energy levels. However, the symmetry of the system also allows us to view the two pairs of CH bonds as two identical oscillators, beeing coupled physically by the CC bond. As a result, the motion of the system will break down into 2 modes, one at the initial frequency of the isolated oscillators, the other slightly blue-shifted. Normal mode analysis places them at 1388 and 1412.5 cm$^{-1}$, to the accuracy of the present simulation. Using (1) with $\nu_{a}=\nu_{0}=1388$ and $\nu_{b}\sim\nu_{0}+v=1412.5$, we find $v=$24.5 cm$^{-1}$=3 meV. Also, the coupling translates into energy swinging between the 2 oscillators with frequency $v$: this is best seen by monitoring both angles after one of them has been changed slightly from its equilibrium value: they vary sinusoidally and out of phase with the expected periodicity, 1.4 ps. If $\theta_{1}$ and $\theta_{2}$ be the angles HCH, then, by symmetry, the amplitudes of the sym and a-s modes are simply $\Sigma=(\theta_{1}+\theta_{2})/2$ and $\Delta=(\theta_{1}-\theta_{2})/2$. At low energy contents these quantities are perfect sinusoids with frequencies $\nu_{a}$ and $\nu_{b}$, respectively; at higher contents, they are affected by amplitude modulation, whose frequency and depth obey (4) and (5). These relations will be used    to study the changes in, and exchange between, the 2 modes with energy content.

\section{Case study: C$_{2}$H$_{4}$, energizing a single vibrational mode}

 As an example, take as target mode the symmetric scissor mode. Normal mode analysis (provided by the semi-empirical codes) determines its wave number to be 1388 to the accuracy of this simulation. A periodic perturbation with $T=0.025$ ps (1333 cm$^{-1}$), making $\delta_{0}=-4\%$, was applied by setting the angles $\theta$ between each pair of C-H bonds at 110 deg, from their equilibrium value, 114.6 deg: each such perturbation induces a potential energy increase of 0.9 kcal/mol or $\sim$0.04 eV. Figure 4 shows the energizing curve: the excitation energy levels off near 110 kcal/mol, a limit imposed by the fact that the red-shifted proper frequency of the free motion has now fallen to 1335 cm$^{-1}$, very near the excitation frequency so the energy gain per cycle tends to zero (see Sec. 3). The peak angular excursion is also limited by stereoconstraints when the angles approach 180 deg: downward distorsion is indeed seen at the crest of the $\theta$ (HCH) sinusoids. 

\begin{figure}
\resizebox{\hsize}{!}{\includegraphics{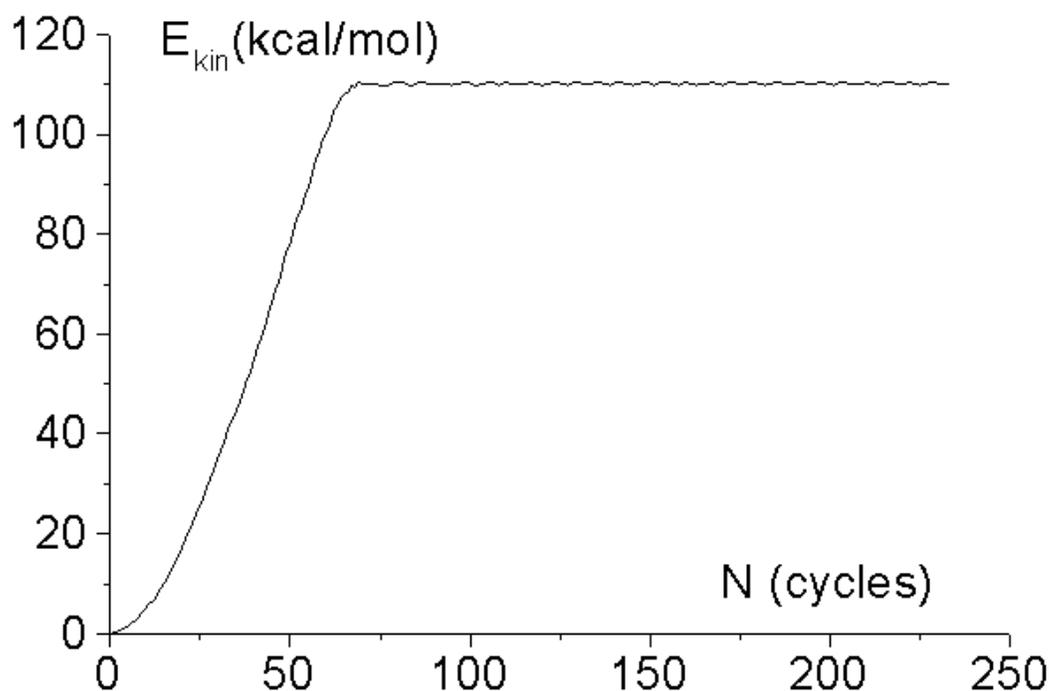}}
\caption[]{Energizing curve for the symmetric scissor mode of C$_{2}$H$_{4}$. The periodic perturbation consists in imposing the same value, 110 deg, on both angles HCH, every 0.025 ps .} 
\end{figure}

The excitation procedure was then repeated, starting from equilibrium but applying only 10 cycles to deliver 5 kcal/mol or about one mode quantum. The molecule was then left to execute its free motion in vacuum. No further evolution is visible for at least 100 ps: on the screen, the molecule is observed to oscillate in an apparently pure symmetric scissor mode with constant amplitude. During this period, the relative energy content of the various modes is monitored in more detail by performing a FFT on the recorded total kinetic (or potential) energy. It is found indeed that the spectrum is dominated by a line very near the nominal target mode frequency. Its harmonics are more than a hundred fold weaker, indicating that the corresponding potential surface is very nearly parabolic. As expected in such circumstances, the mode is only very weakly coupled to other modes and none shows up in the spectrum above a relative level of $10^{-2}$, except the i.p. symmetric CH bending (rock) mode at 835 cm$^{-1}$ and intensity $\epsilon=2\,10^{-2}$ relative to the scissor mode. A procedure to efficiently excite nearly single normal modes of vibration is thus demonstrated.

At the same time, the coupling energies with other modes can be estimated using Sec.2. The frequency separation of the bending mode from the target mode, 555 cm$^{-1}$, being so wide, one can assume that it is much larger than $v$ in (5), so that the latter is approximately 10 cm$^{-1}$ or 1.2 meV. This much is due to the geometrical similarity of the two modes. By contrast, the very close a-s scissor mode is not even detected because it holds a negative phase relation to the symmetric mode. Notwithstanding, it can be excited by increasing the energy content of the sym mode, but what happens then will best be understood after the following experiment is described.

\section{C$_{2}$H$_{4}$: anharmonicity, frequency dragging, mode locking and localization}

\begin{figure}
\resizebox{\hsize}{!}{\includegraphics{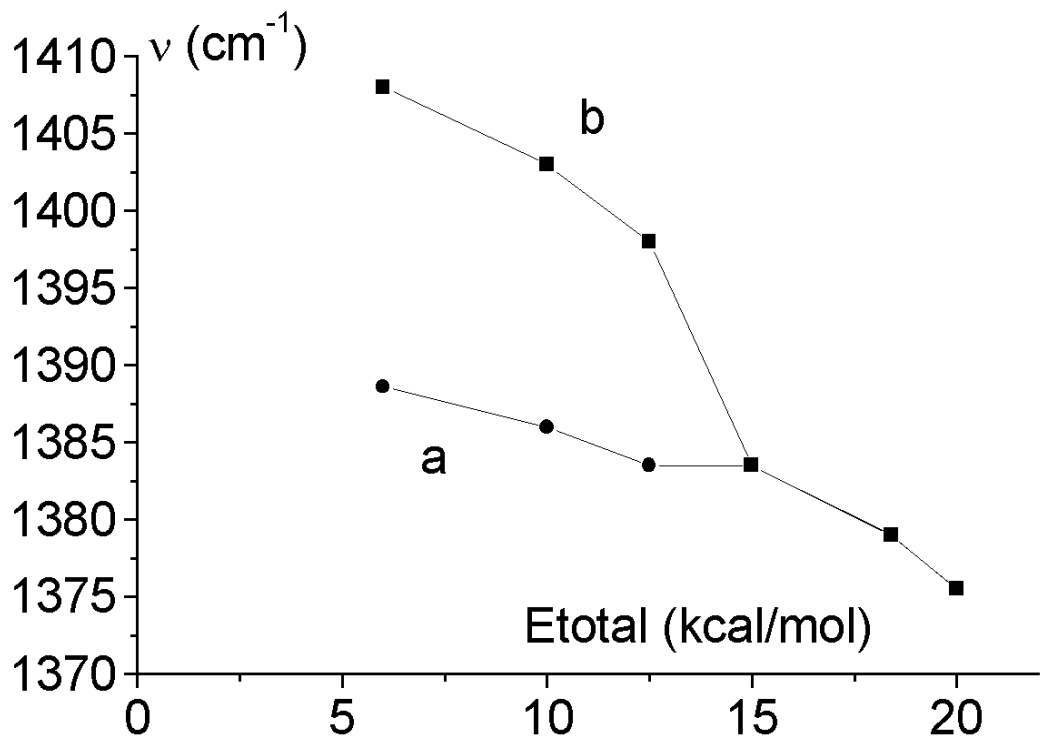}}
\caption[]{Illusrating mode freqency dragging. $a$ and $b$ represent the wave numbers of the symmetric and a-s modes, respectively, as a function of molecular energy content. Note the merging of the two curves at the higher energies, a phenomenon due to anharmonicity alone.} 
\end{figure}

\begin{figure}
\resizebox{\hsize}{!}{\includegraphics{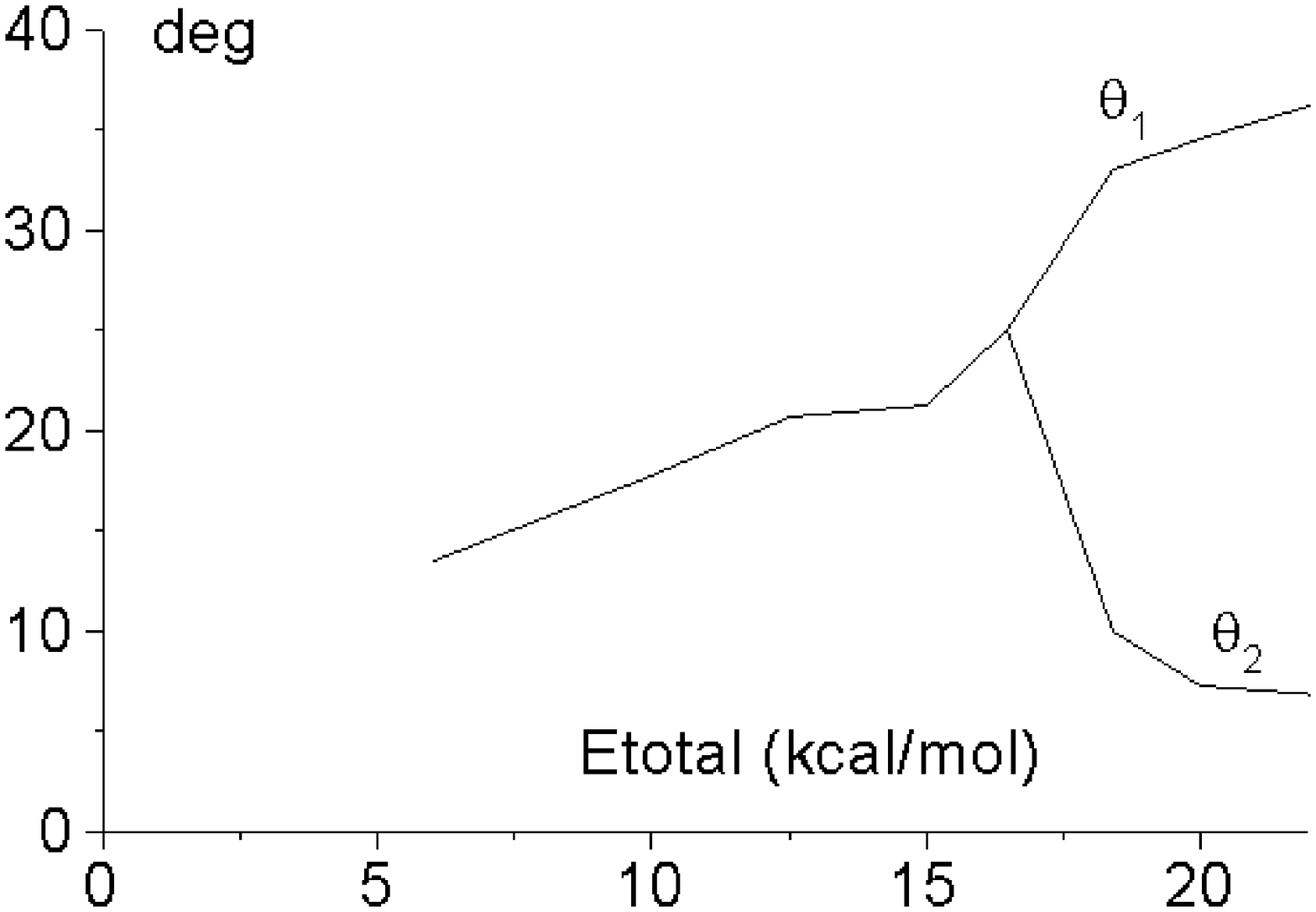}}
\caption[]{$\theta_{1}$ and $\theta_{2}$ represent the maximum angular excursions of the two HCH angles. Note the quenching of oscillator 2 by oscillator 1 at the higher energies.} 
\end{figure}

 Here, the periodic perturbation ($T=0.025$ ps) is not applied to both CH bond angles, as in Sec. 4, but to one of them only, No 1 (asymmetric excitation). As a result of their coupling, oscillator No 2 is also excited and they share the deposited energy, so that both scissor modes, $\Sigma,\Delta$, are indeed excited (see beginning of Sec. 4). We followed their frequencies (fig. 5) as a function of the energy content of the molecule (varied by increasing the number of energizing cycles; see Sec. 3). It is observed that the anharmonic shift is initially small, then increases abruptly when the energy content nears 3 mode quanta. Ultimately, both frequencies merge when the energy exceeds 4  quanta : this demonstrates the phenomenon of frequency dragging and mode locking (see Minorsky\cite{min}). All the while, both amplitudes increase almost in the same proportion (although saturation is ultimately visible on the strongest mode). The behaviour of $\theta_{1,2}$ illustrate still another molecular feature, namely localization (fig. 6): at low deposited energies, the dynamics is linear and both resonant oscillators share the energy equally but alternatively, according to Sec. 2; after an identical initial increase of both peak angles with input energy, the oscillator which was not initially excited starts to weaken and is ultimately completely quenched and the motion becomes confined to that angle which was initially perturbed. All these  phenomena are continuous and linked with anharmonicity and saturation, and do not imply higher-order coupling terms (see end of Sec. 2).

\section{C$_{2}$H$_{4}$: locked ``relaxation" oscillations}

Returning to the symmetric excitation of Sec. 5, we study the effects of increasing the energy content of the sym mode by the same procedure. There appears to be a threshold for the latter between 6 and 7 kcal/mol, beyond which, periodically, the a-s mode develops exponentially then dies out similarly. The turn occurs when the sym mode has provided so much of its energy that it falls below the threshold and therefore can no longer sustain the growth of the a-s mode (fig. 7). As behoves such relaxation mechanisms, the time constant of these evolutions decreases as the energy 
content increases. However, stereoconstraints (Sec. 5) progressively hinder the growth of the weaker (a-s) mode, so much as to finally quench it completely, as was the case in Sec. 6.

\begin{figure}
\resizebox{\hsize}{!}{\includegraphics{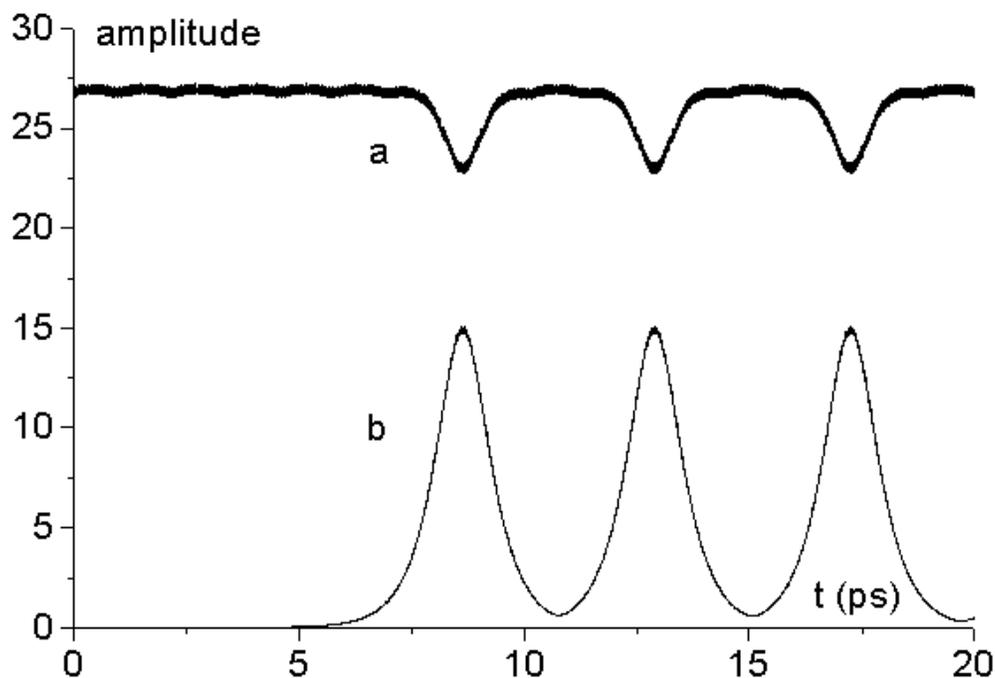}}
\caption[]{Mode locking and relaxation oscillations. With more or less delay following strong initial excitation of the sym scissor mode alone, the a-s mode is energized in bursts of decreasing amplitude as the energy content increases. The a-s mode wave number remains locked to that of the sym mode (see spectrum, fig. 8). Both behaviours are distinctly at variance with that of linear oscillators illustrated in fig. 1. At the limiting level of the energizing curve in fig. 4, the a-s mode is completely quenched.} 
\end{figure}

\begin{figure}
\resizebox{\hsize}{!}{\includegraphics{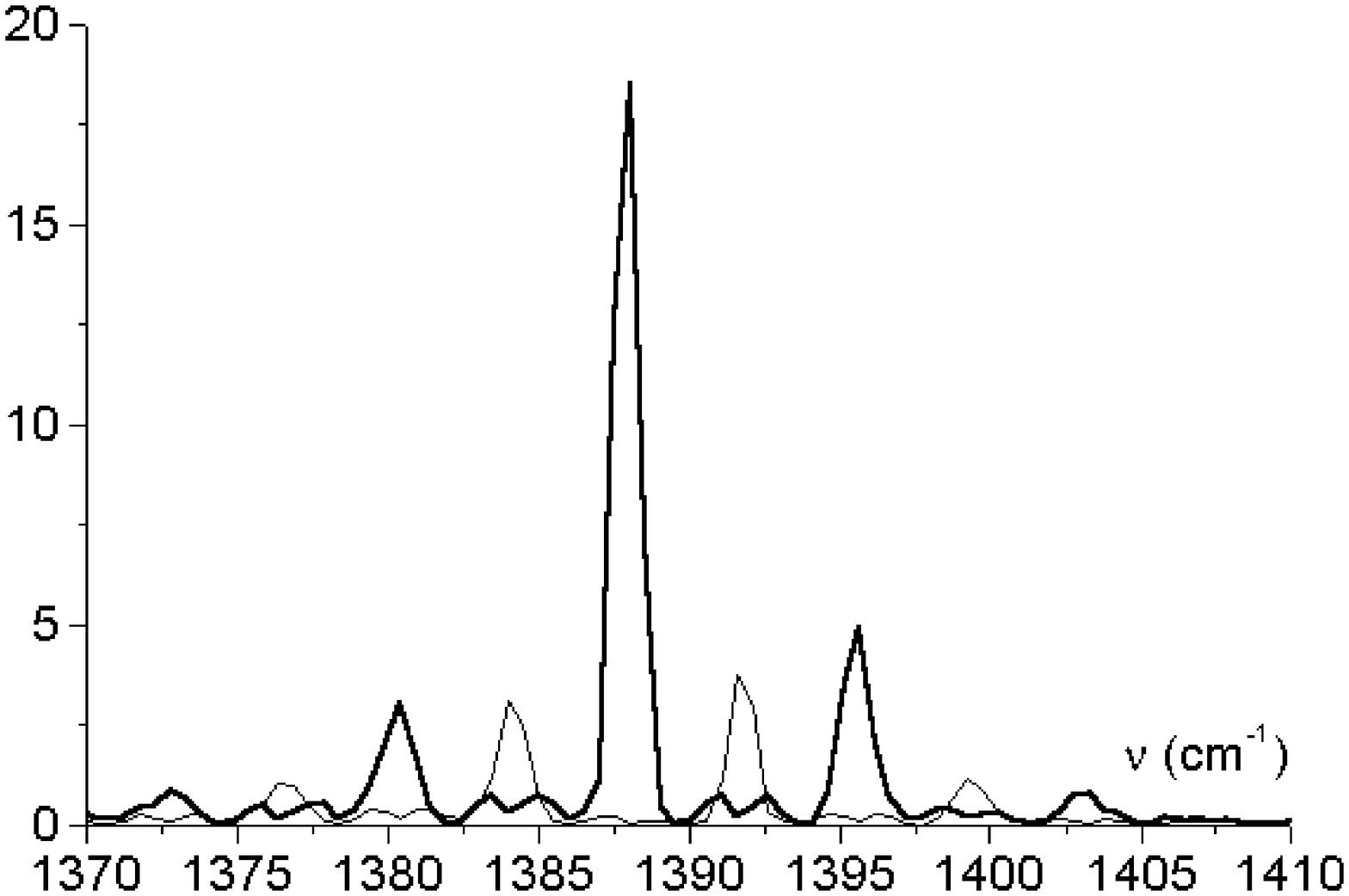}}
\caption[]{Spectra of the angular variation of $\Sigma$ and $\Delta$, the sym (heavy line) and a-s (light line) scissor modes; see Sec. 8 for interpretation.} 
\end{figure}

Obviously, the coupling between these two modes is not of the ``linear" type considered in Sec. 2.
 Numerical solutions on that basis, even including the first-order non-linear corrections, never exhibit the excitation of the a-s mode when, initially, only the sym mode is excited. That is, unless some sort of asymmetry is introduced, even though very weak, such as a slight inequality in $\nu_{1,2}$ or $\theta_{1,2}$. One is therefore led to conclude that the growth of the a-s mode is made possible only by the unstable character of the exciting mode at sufficiently high energy content. Such instabilities are known from non-linear mechanics to occur at the crossings of otherwise stable trajectories (see Wu\cite{wu}). At these points, the trajectories are distorted so as to form two new separate trajectories, each connecting two different parts of the initial trajectories, and  covering different energy domains. In molecular physics this is the well-known phenomenon of potential surface avoided crossing and repulsion, yielding a lower- and a higher-energy surface (see Herzberg\cite{herz}). In the present experiment, we observe the swinging of the system between the two dips of the new lower-energy surface, which still essentially define the characters of the sym and a-s modes. The threshold energy for this to happen is obviously the energy at the tip of the intermediate hump between the dips.

Beyond that, we expect both modes to have the same frequency, progressively red-shifted by anharmonicity as the energy content is increased. This is borne out by the spectrum of fig. 8, for instance, which corresponds to an energy content of 15 kcal/mol. The main peak of $\Sigma$ implies that this mode oscillates at 1388 cm$^{-1}$ and its 2 nearest side-bands, that it is modulated with frequency 8 cm$^{-1}$ ( period 4.3 ps) as evidenced by fig. 7. As to $\Delta$, its 2 main peaks are symmetric with respect to the main peak of $\Sigma$, implying that this mode has the same frequency and the separation of the 2 peaks implies that it is also modulated, and with the same periodicity (cf. fig. 7). These frequency relations are maintained as the energy content varies: when the a-s mode is observable, it is locked to the symmetric mode. 

This behaviour cannot be emulated by coupled-mode theory unless a non-linear correction term in 
$(x_{1}-x_{2})^2$ is included. Such a term makes the system non-integrable, i.e. unstable. It is also responsible for the side-bands apparent in fig. 8. Indeed, the rate of decrease of these side-bands with distance from the center frequency is a measure of the time constant of the exponential rise and fall of the mode amplitudes.

\section{Higher-order resonances}

The next experiment is intended to illustrate ``non-linear" (or ``parametric") resonances, which give rise to combination frequencies and Fermi resonance. Here, the target mode is the sym CH stretch. An equal extension of 1.15 \AA{\ } (from the equilibrium value, 1.1 \AA{\ }) is applied periodically, every 0.011 ps (3000 cm$^{-1}$, $\delta_{0}=5\%$) to all 4 C-H bonds. After 10 cycles, the acquired excess energy is 9 kcal/mol (0.39 eV), nearly the energy of 1 mode phonon. The free motion of the molecule is monitored thereafter.

\begin{figure}
\resizebox{\hsize}{!}{\includegraphics{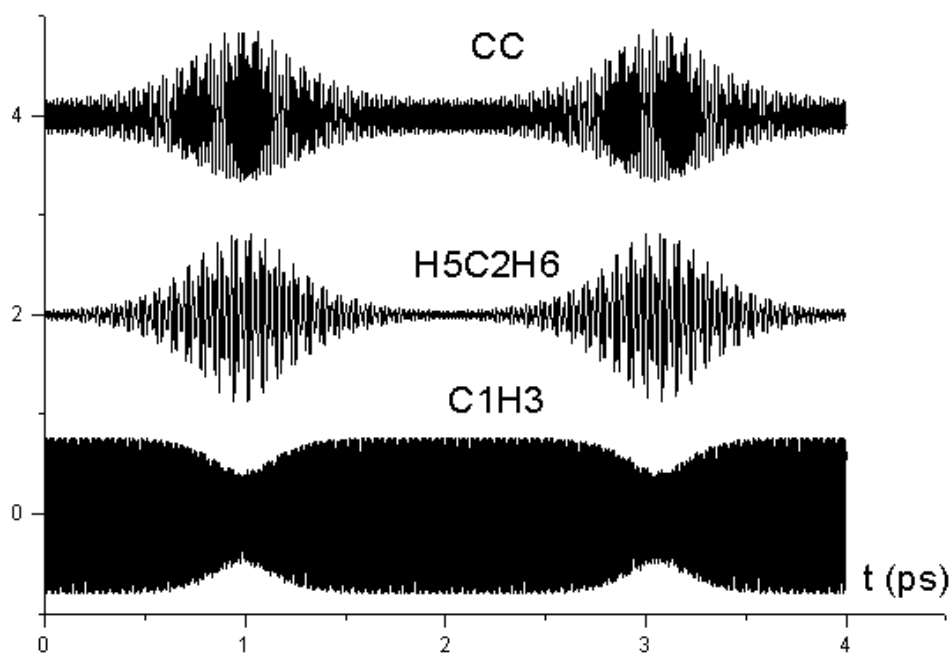}}
\caption[]{Higher-order resonance. The symmetric CH stretch mode $\nu_{1}$ is initially excited. Immediately thereafter, some of its energy begins to be transferred periodically to both the CC stretch, $\nu_{2}$, and sym scissor, $\nu_{3}$, modes, all 3 modes being of same symmetry species $A_{g}$. Again, the envelope shapes witness to the non-linear character of the coupling giving rise to a 2nd-order resonance  $\nu_{1}=\nu_{2}+\nu_{3}$.  For about 10 ps, the spectra of these motions exhibit only these 3 frequencies.} 
\end{figure}

Almost immediately, the angles HCH and the bond length C-C are set in oscillation together with the C-H bond lengths (fig. 9). The former experience periodic surges in amplitude, for $\sim0.5$ ps, every 2.1 ps, perfectly in tune with a strong decrease of the CH bond length amplitude of vibration. This suggests a very non- linear regime with energy swinging between the C-H motion, on the one hand, and the HCH and C-C motions on the other hand. The spectrum of each type of motion is indeed very pure , peaking respectively at 3239-3243, 1400-1403 and 1839-1840 cm$^{-1}$. Obviously, the former is the sum of the latter two: this resonance occurs only because the molecular energy content is now high enough to perturb the parabolic shape of the potential energy surfaces to the point of inducing a significant higher term in the mode coupling terms (see end of Sec. 2). Because of the periodic energy exchange between modes, each band is accompanied by side-bands correspondig to sum and difference frequencies. Note that, due to imperfections of the AM1 type of chemical modeling used here, the wave numbers differ slightly from experimental ones, without affecting the phenomenological behaviour.

Further analysis of the bond angle and length variations reveals that the first band corresponds to  pure symmetric C-H stretch (all 4 bonds being sychronized) and the second to pure symmetric scissor, to the accuracy of the present simulation.

After about 10 ps, the motion becomes more complex: fig. 10 shows the FFT of the total molecular kinetic energy. In more detail, the 2 pairs of C-H stretches are no longer synchronized, neither are the 2 HCH angles. In order to break their motions down into modes, we take the spectra of (C$_{1}$H$_{3}$+C$_{1}$H$_{4}$)$\pm$(C$_{2}$H$_{5}$+C$_{2}$H$_{6}$) and H$_{3}$C$_{1}$H$_{4}$$\pm$H$_{5}$C$_{2}$H$_{6}$. We now have 2 bands for each of these combinations: respectively

sym CH str 3210, 3245 cm$^{-1}$, 

a-s CH str 3220, 3264 cm$^{-1}$, 

sym scissor 1377, 1413 cm$^{-1}$, 

a-s scissor 1389, 1432 cm$^{-1}$.

\begin{figure}
\resizebox{\hsize}{!}{\includegraphics{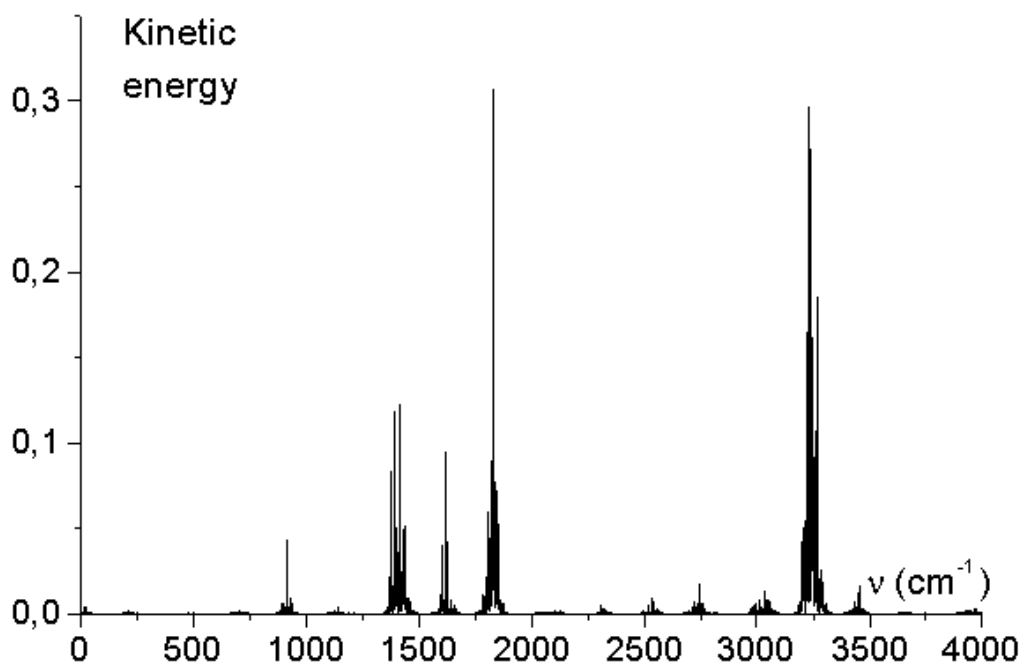}}
\caption[]{Follow-up to fig. 9. About 10 ps after energizing the CH stretch mode, the spectra become more complicated. The spectrum of the total molecular kinetic energy represents here the new (and final) energy distribution between modes, in kcal/mol/cm$^{-1}$. The initial sym CH str energy is now partitioned between this mode and the a-s CH str mode (58 $\%$), the CC str mode (21 $\%$) and the 2 scissor modes (20 $\%$), totalling 99 $\%$ between them.} 
\end{figure}

The CC str is only slightly shifted to 1832 cm$^{-1}$. Each of the 4 CH str wave numbers is found to be the sum of the wave numbers of the CC str and one of the two scissors. Thus, the same type of non-linear resonance now extends to the scissor and CH str modes (of course, the 2 other CH str modes can hardly couple with the scissors to generate a resonance). 

Here again, the qualitative aspects of these observations can only be mimicked, using the equations of Sec.2, if the term $(x_{1}-x_{2})^2$ is included, which allows the formation of combination frequencies. The other term, affecting explicitely the oscillators' resonant frequencies, brings no qualitative change to the picture. A quantitative determination, by cut-and-try, of the perturbative terms responsible for specific observations is  outside the scope of the present work

Finally, after about 100 ps, steady-state seems to be established, and the relative amounts of energy in the scissor, CC str and CH str types of motion are 0.34, 0.36 and 1, respectively; the rest is negligible. Thus, only 5 modes out of 12 have been excited, and then, not even equally. A similar conclusion can be drawn from the selective excitations of other modes. For instance, if one mode quantum is deposited in the CC stretching vibration, less than 1$\%$ is transferred to the other modes.

\section{Transition to irreversible IVR}
\begin{figure}
\resizebox{\hsize}{!}{\includegraphics{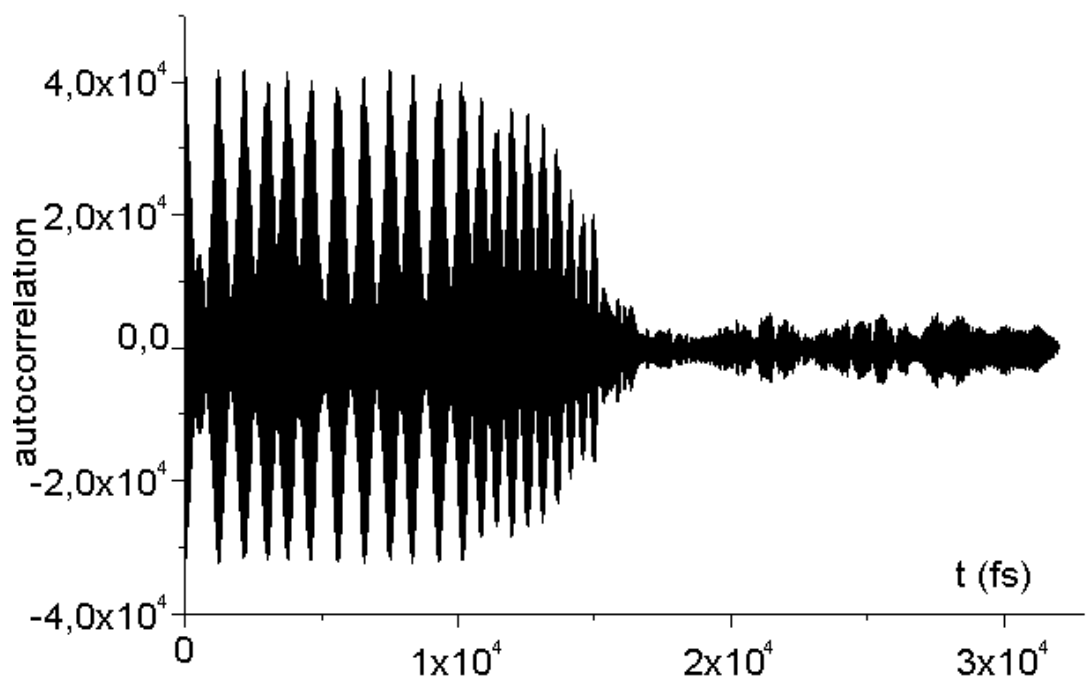}}
\caption[]{Illustrating irreversible transition from a regular motion to a partially chaotic regime, following the stretching of a C-H bond in ethylene (see Sec. 9)} 
\end{figure}

   When the energy deposited locally or in a single mode becomes large, new phenomena are observed. In the following experiment, one of the C-H bonds was stretched to 1.2 \AA{\ }, giving the molecule an excess energy of 20 kcal/mol and left to move freely. Figure 11 shows the autocorrelation of the total kinetic energy as a function of time. The first picoseconds are characteristic of regular (quasi- harmonic) motion: note the strong periodic peaks with a weak and slower amplitude modulation superimposed, indicating the dominance of CH str oscillations. Soon, a steep fall in the peak amplitude becomes apparent and the pattern changes condiderably: the correlation is substantially reduced and it becomes difficult to recognize any regularity in the modulation of its envelope. At the same time, many modes other than the stretches show up in the spectra. This experiment illustrates an irreversible path to chaos, namely increasing energy density. However, the deposited energy here is soon diluted, even in so small a molecule, and the irreversible process comes to a stop.

\section{Large molecules}

Another favorable factor for chaos is the size of the oscillating system, because this increases the spectral density of modes, and thus favours the overlap of resonances (Ford\cite{ford}, Sibert \etal\cite{sib}) and opposes the spatial dilution of energy. In order to illustrate this phenomenon, the same type of computational experiments were performed on a much larger hydrocarbon molecule, composed of interlinked peri- and catacondensed carbon rings, and including benzene, pyrene and pentagonal rings, 8 oxygen, 2 sulphur and 1 nitrogen atoms, 11 tertiary, 44 secondary and 9 primary H atoms for a total of 101 atoms, fully described in Papoular\cite{pap}. The structure is 3-D and has linear extensions of order 1 nm. The lengths and relative orientations of bonds are initially computationally optimized, at rest, for a minimum total potential energy.

The large number of normal modes in this case (297) uncovers new interesting behaviour, which extends over longer periods of time. The computational simulation then is no longer feasible at the higher-quality level, so a less sophisticated software was used here: Molecular Mechanics MM2. While not affecting severely the general features of the underlying physics, this is very much faster and allowed the extension of simulations up to 1 ns (Papoular\cite{pap}).

\begin{figure}
\resizebox{\hsize}{!}{\includegraphics{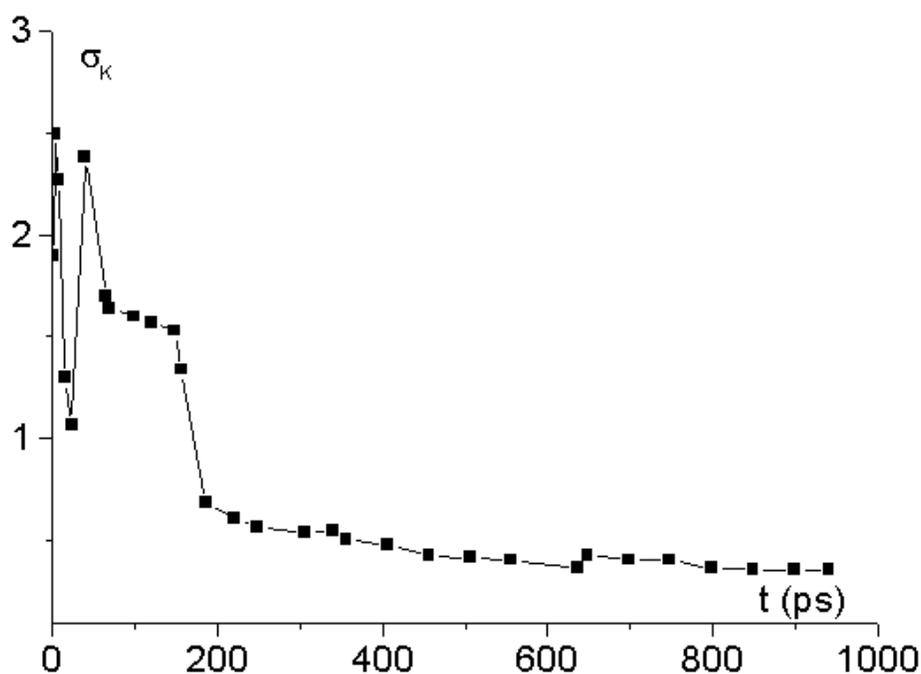}}
\caption[]{Relaxation of a 101-atom hydrocarbon molecule following selective excitation of CH sretchings, illustrated by the evolution of the standard deviation of the total kinetic energy (kcal/mol). Note the abrupt downward transitions near 50 and 170 ps.} 
\end{figure}

Another difficulty lies in devising a priori a system of simultaneous perturbations of all bonds so as to emulate a prescribed normal mode. Fortunately, this turns out not to be essential for present purposes. In the following experiment, the target modes are the CH stretches. The energizing procedure consists in imposing periodically ( $T=0.012$ ps, $\nu=2800$ cm$^{-1}$) to 2 CH bonds at opposite edges of the molecule (one aliphatic and one aromatic) an extension of 1.15 \AA{\ }, from their equilibrium value, $\sim$1 \AA{\ }. Twenty cycles are enough to deposit 8.5 kcal/mol ($\sim$1 mode quantum) in the molecule.

\begin{figure}
\resizebox{\hsize}{!}{\includegraphics{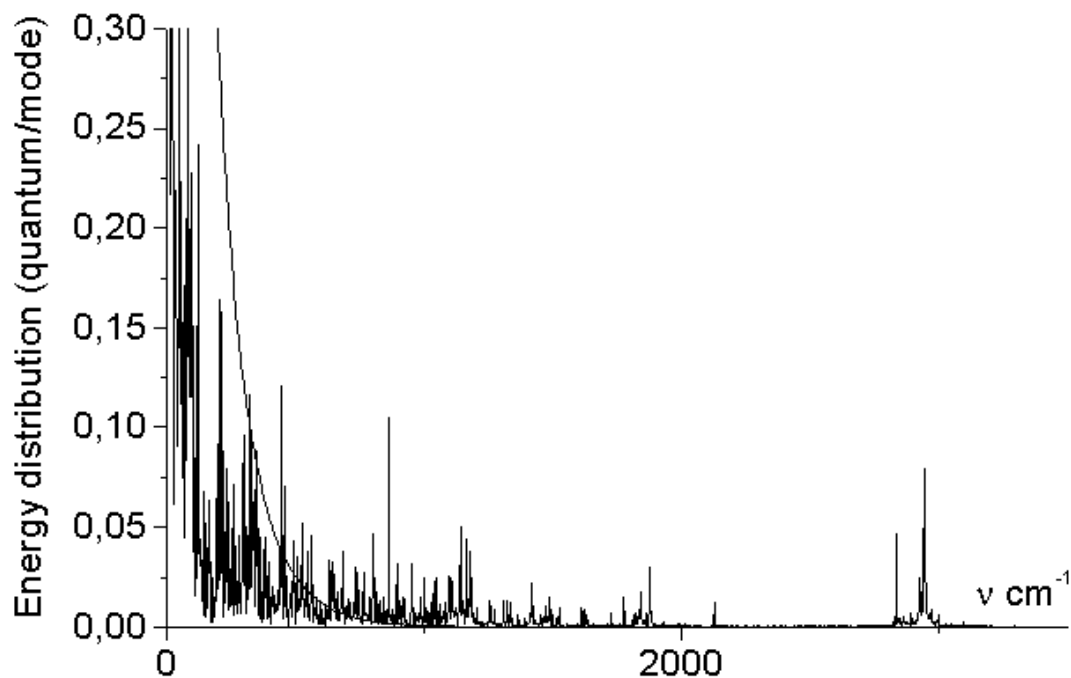}}
\caption[]{Average number of mode quanta, $h\nu$, per mode $\nu$, (solid line), deduced from FFT of the total kinetic energy over the final 8 ps of the dynamics, for the case of fig. 12. The dashed line represents the thermodynamic equilibrium at temperature 130 cm$^{-1}$, computed for the same energy content, 2900 cm$^{-1}$.} 
\end{figure}

The following free-motion relaxation is not smooth. Its different phases are strikingly illustrated by monitoring the standard deviation, $\sigma_{K}$, of the total kinetic energy, $E_{K}$: statistical arguments show that, roughly, $\sigma_{K}/E_{K}=n^{-1/2}$, where $n$ is an effective number of vibrational modes, considered as gas particles (see Reif\cite{reif}). Figure 12 shows $\sigma_{K}$ (computed over 8 ps for each point) as a function of time. This uncovers 3 different phases separated by ``abrupt" transitions and differing in the spectral distribution of energy:

a) 0 to 50 ps. The energy remains localized mainly in the two perturbed bonds; the spectrum of $E_{K}$ exhibits only their two different local frequencies, 2775 and 2809 cm$^{-1}$, still very close to the excitation frequency. This spectral sparseness is the cause of the large fluctuations in $\sigma_{K}$.  

b) 50 to 160 ps. $\sigma_{K}$ now decreases smoothly because many other C-H bonds have been strongly excited. They are mainly aliphatic and oscillate at $\sim$2820 cm$^{-1}$, redward of their natural frequency because of their high energy content. Some energy has gone into in-plane CH bendings and CC stretches. 

c) 160 to 1000 ps. In this final phase, delocalization is complete, and both aliphatic and aromatic CH str are found near their natural frequencies, 2835 and  2940 cm$^{-1}$ (to the accuracy of this software). Most importantly, lower frequency modes are progressively and smoothly excited down to the lowest. Figure 13 shows the spectral energy distribution during the last 8 ps, in terms of average number of mode quanta per mode.

We take this to be steady-state distribution of energy because fig. 12 shows that $\sigma_{K}$ is close to its asyptotic value, estimated at 0.36. This value implies that $\sim$140, out of the total 297, modes have been effectively excited, most of them at the lower end of the spectrum, mainly skeleton modes. Of the energy initially deposited in the molecule, 28 $\%$ are still in the target CH str modes. It must be stressed that these are average values for a steady but dynamic state. Energy is constantly exchanged between $\emph{modes}$, and this corresponds, in quantum mechanics, to the flow of energy throughout all the molecular $\emph{states}$ of the portion of phase space that is available at this energy content, taking into account selection rules and all other restrictions. If the 297 quantum coupled oscillators had somehow settled down into a statistical thermodynamic equilibrium with the same energy content, its temperature would have reached 130 cm$^{-1}$ (see Allamandola \etal\cite{atb}), and the average number of mode quanta in mode $\nu$ would be $1/[exp(\nu/130)-1]$, as illustrated by the dashed curve in fig. 13. The difference is striking and due to the much larger phase space covered in a fully ergodic motion, subject only to energy conservation, which  greatly favours low frequencies and penalizes high frequencies.

\begin{figure}
\resizebox{\hsize}{!}{\includegraphics{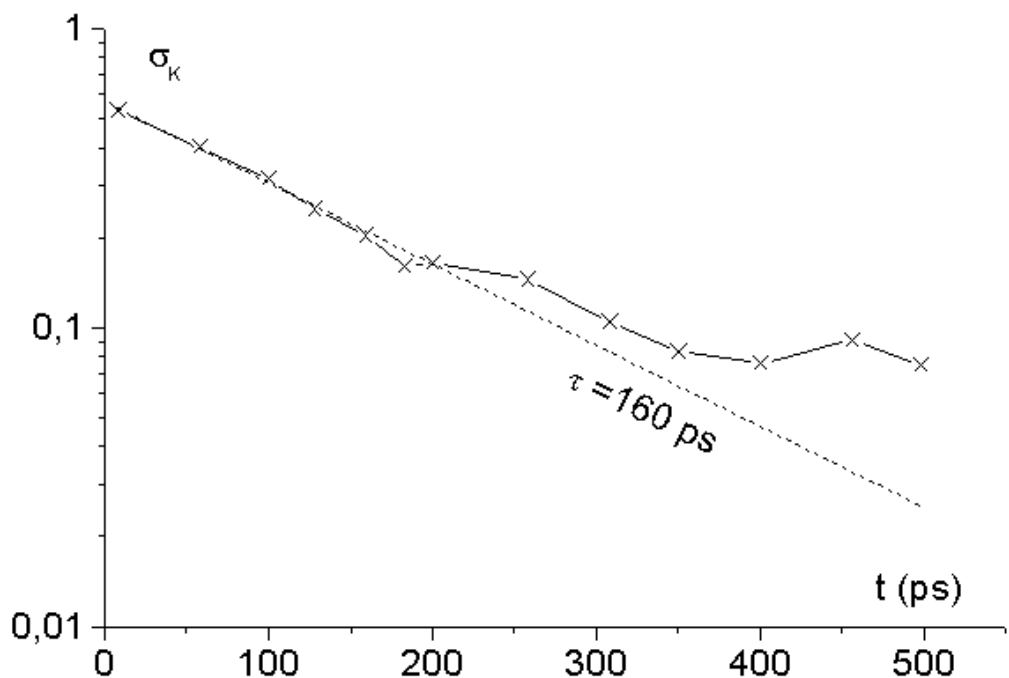}}
\caption[]{Same as fig. 12, following deposition of 810 cm$^{-1}$ in a CC str mode. Here, the vertical scale is logarithmic so as to highlight the initial exponantial character of the relaxation.} 
\end{figure}

\begin{figure}
\resizebox{\hsize}{!}{\includegraphics{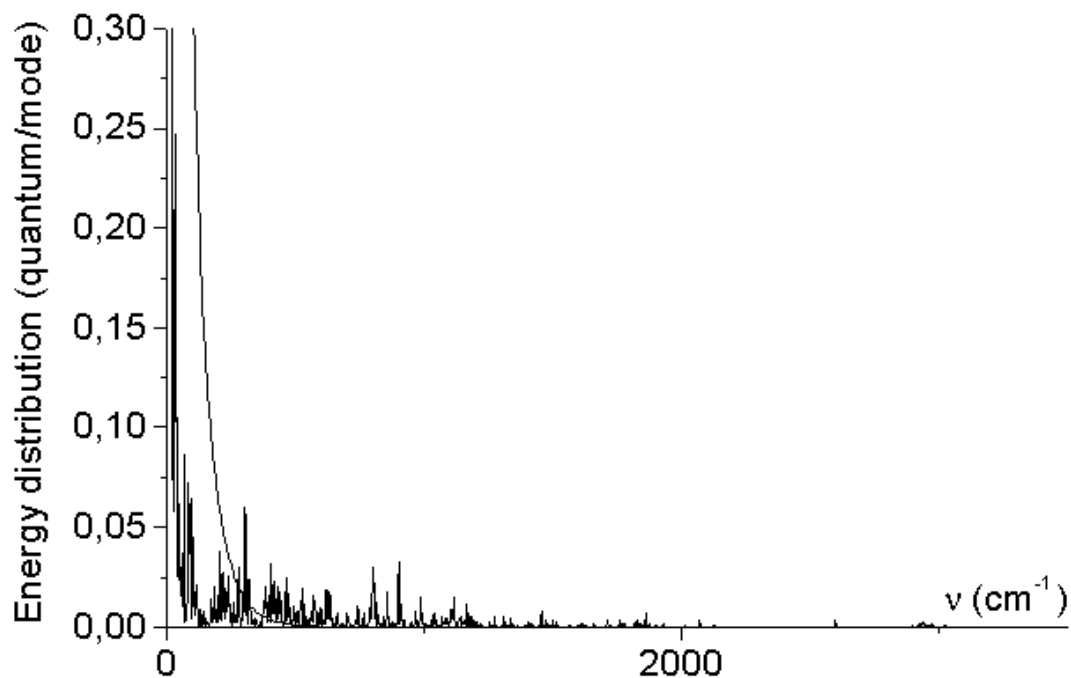}}
\caption[]{Same as fig. 13, for the case of fig. 14. The dashed line is for thermodynamic equilibrium at $T=70$ cm$^{-1}$, corresponding to the same energy content, 810 cm$^{-1}$.} 
\end{figure}

The total relaxation time is about 1 ns. By contrast, the transition time is only of order 10 ps. During this time, larger and larger parts of the molecule are seen to be quickly set in motion.  This signals again the existence, at high local energy densities, of metastable states which soon lead to significant redistribution of energy among modes and changes in the spectra.

Different types of relaxation are encountered when initially energizing different normal modes. In the following and last experiment, one of the CC bridges between aromatic ring clusters was energized by periodic perturbations of its length, with frequency 926 cm$^{-1}$. Again, the frequency selectivity of the process makes for a very sparse spectrum in the first 10 ps of the free motion: only 2 modes are excited, at 876 and 948 cm$^{-1}$, which oscillate coherently (i.e. exchange energy periodically, as in Sec. 2), giving rise to beatings in the total kinetic and potential energies, with period 0.225 ps, corresponding precisely to their frequency difference. However, the relaxation process immediately sets in. It is now exponential for about 200 ps, with an exponent of 135 ps (fig. 14). Afterwards, it begins to level off but still exhibits long period fluctuations, also seen in the auto-correlation functions of the total energies. This indicates  bulk structural vibrations at very low frequency, as borne out by the final spectral energy distribution, fig. 15, where the dominance of low  frequencies is even more strongly pronounced than in the previous case.

\section{Conclusion}

Single-mode (or nearly so) excitation of isolated molecules allows one to explore different relaxation paths towards different dynamic steady-states. Along the way, it is possible to indentify bottlenecks and determine coupling strengths between modes. The high temporal resolution of computational modeling uncovers phenomena which are characteristic of the non-linear domain of molecular dynamics, but not very easy to demonstrate experimentally, such as local modes, frequency dragging and mode locking, as well as metastable states (saddle points) and relaxation oscillations.

Coupled-oscillators theory helps understanding the underlying physics, disentangling complicated states and quantifying the parameters responsible for these phenomena.

The final relaxed states of the molecules are found to be different from each other and from thermodynamic equilibrium. Computational experiments may therefore help preparing molecules for specific chemical reactions, or predicting their IR emission following specified excitation, which was the initial goal of this research.

\end{document}